\documentclass{PoS}
\usepackage{mathrsfs}
\usepackage{eucal}
\usepackage{lineno}

\title{Recent results from the VERITAS multimessenger program}

\ShortTitle{The VERITAS multimessenger program}

\author{
\speaker
{Marcos Santander}$^{1}$ for the VERITAS Collaboration\footnote{for collaboration list see PoS(ICRC2019)1177}\\
{\itshape \href{https://veritas.sao.arizona.edu/}{https://veritas.sao.arizona.edu/}}\\
{$^{1}$ \itshape Department of Physics and Astronomy, University of Alabama, Tuscaloosa, AL 35487-0324, USA}\\
E-mail: \email{jmsantander@ua.edu}
}

\abstract{
VERITAS, an array of atmospheric-Cherenkov telescopes sensitive to gamma rays in the very-high-energy range (VHE, E > 100 GeV), carries out an extensive multimessenger program focused on the search for electromagnetic counterparts to high-energy neutrinos and gravitational waves. As both neutrinos and gamma rays are expected to be produced in hadronic interactions near cosmic ray accelerators, the detection of a gamma-ray source in temporal and spatial coincidence with the arrival of astrophysical neutrinos could reveal cosmic ray sources and provide insights into their properties. The detection of gravitational waves by LIGO in coincidence with a gamma-ray burst has opened exciting possibilities for the study of these powerful transients. The detection of VHE gamma rays from these sources would not only help in pinpointing an electromagnetic counterpart to the gravitational-wave event, but also enable the study of the relativistic shocks responsible for the high-energy emission. This talk will present an overview of the VERITAS multimessenger program, highlighting recent results from the search of VHE gamma-ray sources associated with neutrino and gravitational-wave events and discuss future plans for these studies.
}

\FullConference{36th International Cosmic Ray Conference -ICRC2019-\\
		July 24th - August 1st, 2019\\
		Madison, WI, U.S.A.}

\usepackage{amsmath}

\begin{document}

\section{Introduction}

The detection of high-energy astrophysical neutrinos in the TeV-PeV range by IceCube~\cite{Aartsen:2013jdh} and the observation of gravitational waves by LIGO/Virgo opens new opportunities for the study of the high-energy transient universe.

High-energy neutrinos are produced in cosmic ray interactions with ambient matter or radiation fields near their source or during propagation. The pionic production of these neutrinos should be accompanied by a flux of high-energy gamma rays in a similar energy range~\cite{Kelner:2006tc, Kelner:2008ke}. Therefore, it is expected that a joint detection of neutrinos and gamma rays would be a smoking gun signature for the location of a cosmic ray accelerator. As no clustering has been observed in the astrophysical neutrino sky that could reveal a significant source, it is possible to boost the sensitivity of these searches by means of a multimessenger approach that searches for gamma-ray emission at the sky location of neutrino events with a high probability of being astrophysical in origin. For this reason, the VERITAS gamma-ray observatory has implemented a wide-ranging neutrino follow-up program to search for very-high-energy (VHE, E $>$ 100 GeV) gamma rays associated with IceCube neutrino events. 

Gravitational waves have been detected by LIGO/Virgo from the coalescense of binary black hole (BBH)~\cite{Abbott:2016blz} and binary neutron star (BNS) systems. The first detection of a BNS merger, GW170817~\cite{GBM:2017lvd}, was also associated with a gamma-ray burst (GRB170817A) triggered by a kilonova. As gamma-ray bursts (GRBs) have recently been detected in the VHE band by MAGIC (GRB 190114C~\cite{2019ATel12390....1M}) and HESS (GRB 180720B\footnote{\href{http://tevcat.uchicago.edu/?mode=1;id=329}{http://tevcat.uchicago.edu/?mode=1;id=329}}), it is possible that a joint detection of GWs and VHE emission from a GRB could be enabled by rapid follow-up observations of a GW localization region associated with a BNS merger. The sensitivity of current GW detectors defines a horizon for the detection of BNS mergers, currently set at 120-170 Mpc. This benefits a VHE detection, as the high-energy emission would not be strongly attenuated by the extragalactic background light. For reference, the optical depth to 100 GeV photons reaches unity at redshift $z\sim1$.

VERITAS~\cite{Holder:2006gi} is a ground-based instrument for VHE gamma-ray astronomy with maximum sensitivity in the 100 GeV to 30 TeV range. It consists of an array of four 12-m optical telescopes each equipped with a camera containing 499 photomultiplier tubes (PMTs) covering a field of view of 3.5$^{\circ}$. The array is located at the Fred Lawrence Whipple Observatory (FLWO) in southern Arizona (31$^{\circ}$ 40' N, 110$^{\circ}$ 57' W, 1.3 km a.s.l.). The angular resolution of VERITAS is $\sim0.1^{\circ}$ at 1 TeV (for 68\% containment) and the energy resolution is 15-25\% at the same energy. VERITAS can detect a VHE gamma-ray source with a spectrum similar to the Crab nebula in about 2 min at $5\sigma$ significance, and a source with a flux of 5\% of the Crab nebula in 1.3 hours. VERITAS collects about 1300 hours of observations per year.

The VERITAS multimessenger program searches for VHE emission associated with neutrino and gravitational wave events. We here summarize the status and highlight results from the program and outline plans for future studies.

\section{Outline of the Multimessenger Program}\label{sec:followup}

\begin{figure}
    \centering
    \includegraphics[width=0.8\textwidth]{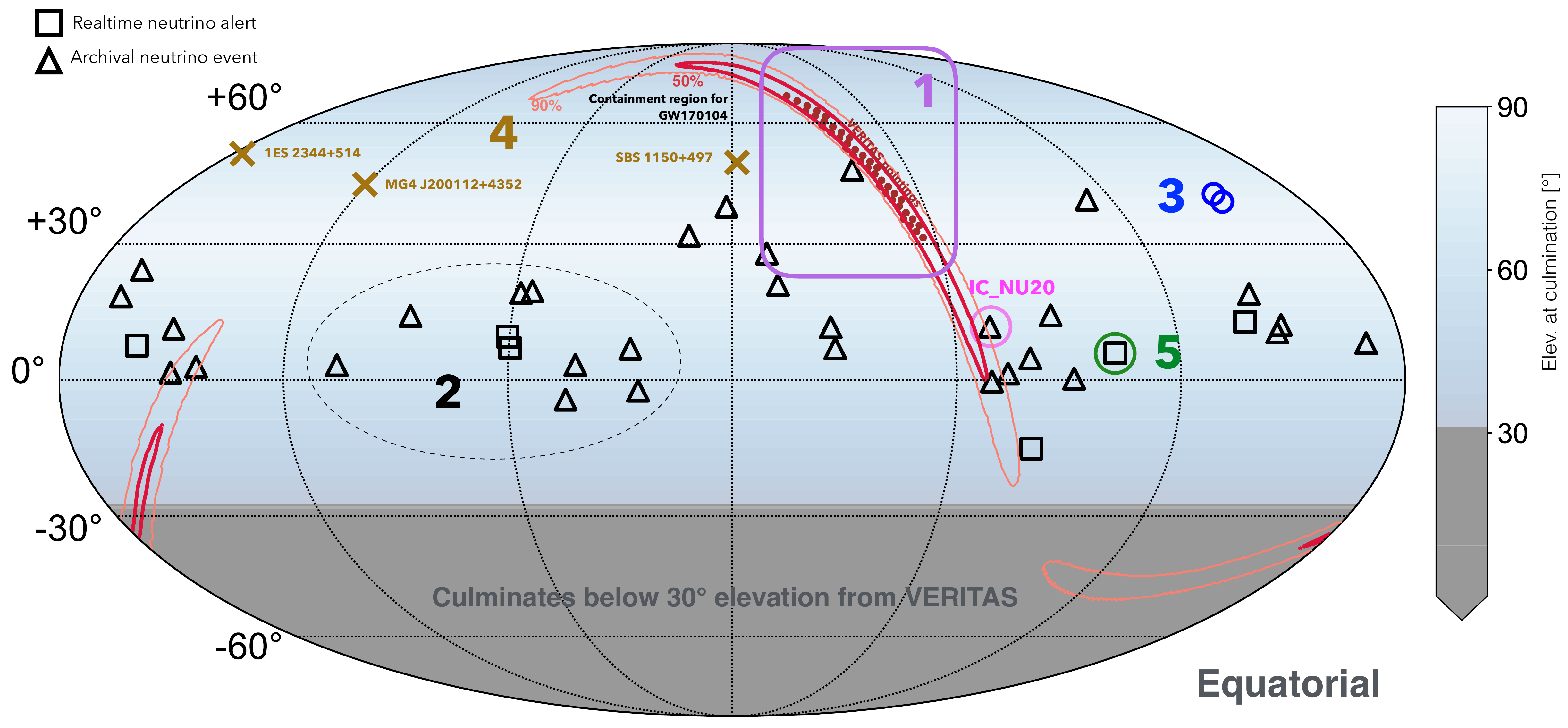}
    \caption{Sky map showing multimessenger triggers observed with VERITAS. The background color indicates the zenith angle at culmination from the VERITAS site for different declination bands. The numbers correspond to the subsection numbers in Sec.~\ref{sec:followup}.}
    \label{fig:allsky}
\end{figure}

The VERITAS multimessenger program~\cite{Santander:2016chw} comprises several prompt follow-up and long-term monitoring activities, which are outlined below and shown in the sky map in Fig.~\ref{fig:allsky}. 

\subsection{Follow-up of gravitational wave events}

\begin{figure}
    \centering
    \includegraphics[width=0.49\textwidth]{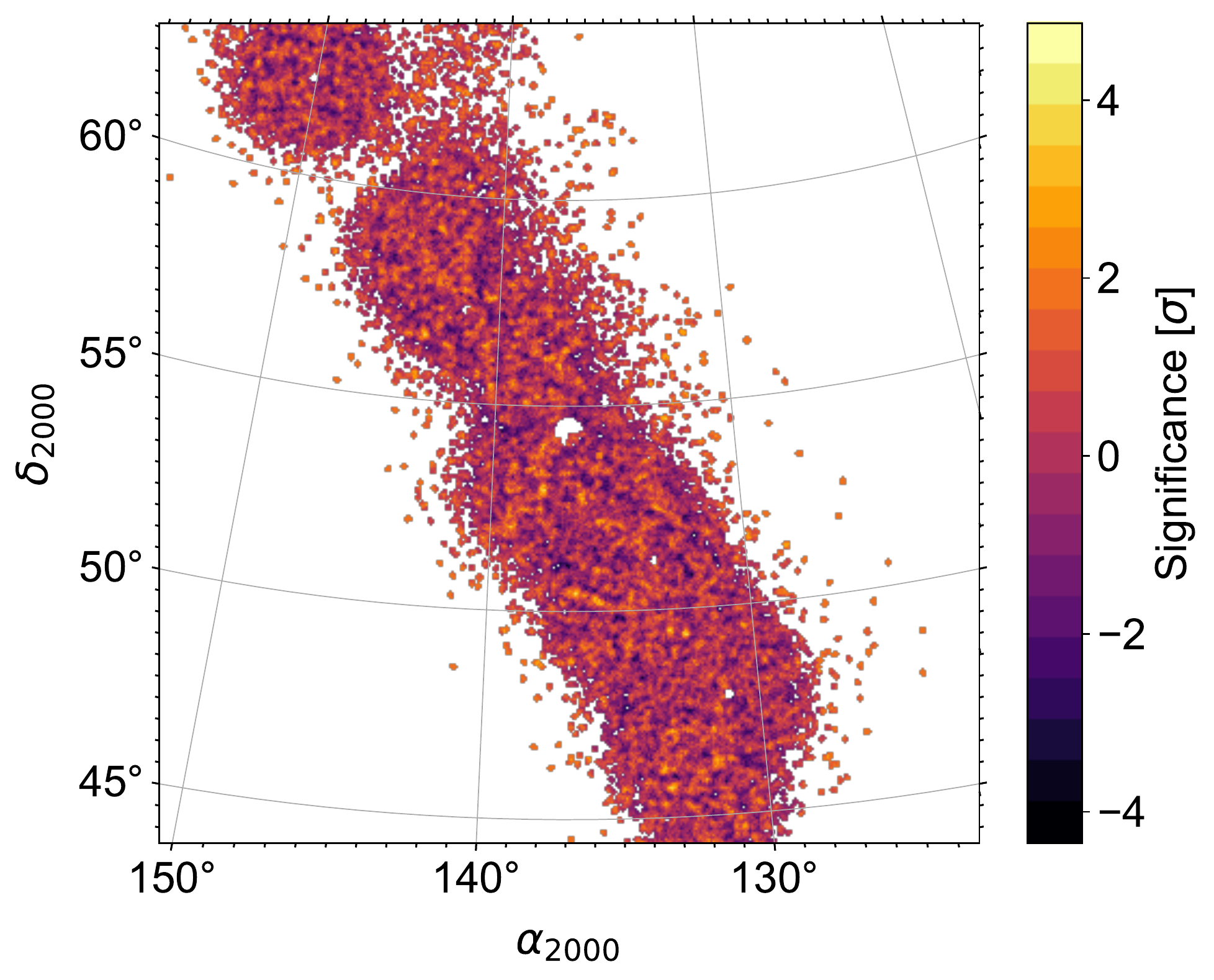}
    \includegraphics[width=0.49\textwidth]{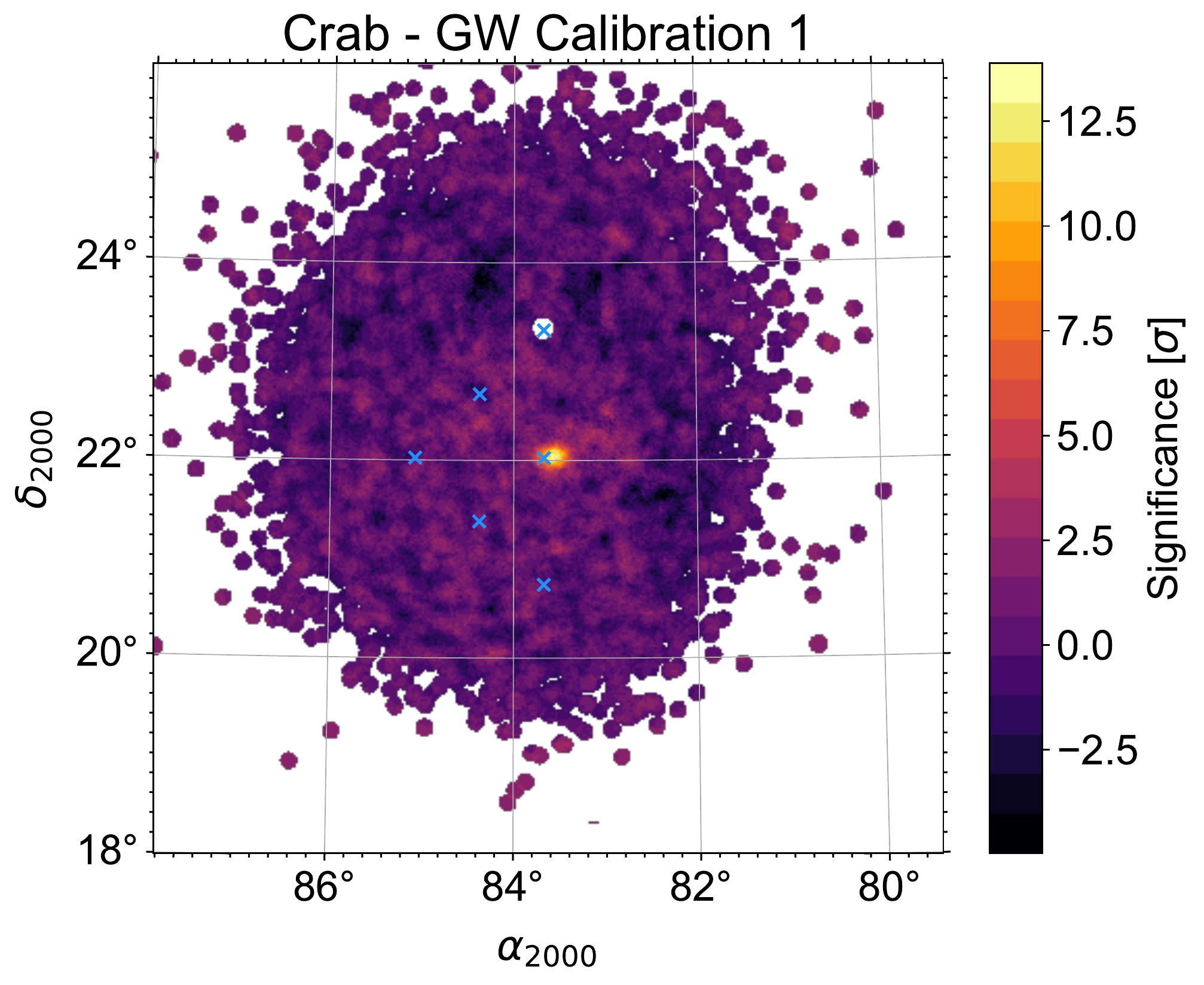}
    \caption{\emph{Left}: Significance sky map for the observations of GW170104. \emph{Right:} Significance sky map for a mini survey with similar pointing offset and 5 min exposure time per pointing around the location of the Crab nebula. The pointings are shown as blue markers.}
    \label{fig:gw170104}
\end{figure}

VERITAS performed the first systematic follow-up of a GW event by an IACT on Jan 4, 2017. The event (GW170104~\cite{Abbott:2017vtc}) was detected during the O2 run of LIGO/Virgo and was triggered by a 50-M$_{\odot}$ binary black hole merger at redshift $z$=0.2. The event and a first map was circulated through the GRB Coordinates Network (GCN\footnote{\href{https://gcn.gsfc.nasa.gov/}{https://gcn.gsfc.nasa.gov/}}) approximately 5 hours after the detection and 39 pointings were calculated to cover the northern section of the 50\% containment region of the event. Pointings were limited to those visible at $>50^{\circ}$ culmination from VERITAS during that night (see Fig.\ref{fig:allsky}). Each pointing was observed for about 5 min, with an average spacing between pointings of 1.83$^{\circ}$, covering 27\% of the containment probability of the event. No evidence for a VHE gamma-ray source was found during this follow up, shown in Fig.~\ref{fig:gw170104}. Under ideal conditions these observations would have been sensitive to sources with a flux of 50\% of the Crab nebula. However, the presence of clouds makes it challenging to set upper limits at this time.\footnote{\href{https://gcn.gsfc.nasa.gov/gcn3/21153.gcn3}{https://gcn.gsfc.nasa.gov/gcn3/21153.gcn3}}

The performance of the tiling method has been evaluated by performing a mini survey of six pointings around the Crab nebula, with the same spacing and exposure time as the observation of GW170104. The goal was to evaluate the behaviour of the background estimation in short exposures and any potential issues introduced by the large source offsets. The results, shown in Fig.~\ref{fig:gw170104}, validate the tiling method with a high-significance detection of the Crab nebula and the correct estimation of the background level.

The current run of LIGO/Virgo, named O3, has resulted in a significant increase in the rate of GW detections. The delay between the GW trigger and the circulation of the first maps has also been reduced from hours to minutes in most cases, which enables fast follow-ups by VERITAS and other IACTs. So far during O3, four GW event follow-ups have been performed out of 18 GW triggers reported by LIGO/Virgo, mostly limited by the visibility of the GW uncertainty region from the VERITAS site at the time of detection. The O3 follow-ups are listed in Table~\ref{tab_gwevents}. 

\begin{table*}
\small
\centering
\begin{tabular}{ccccc}
\hline
\textbf{GW ID} & \textbf{Delay} & \textbf{Compact binary} & \textbf{Prob. covered} & \textbf{VERITAS obs.} \\
 & \textbf{[hrs]} & \textbf{coalescence type} &  &  \textbf{[hrs]} \\
\hline

\href{https://gcn.gsfc.nasa.gov/notices_l/S190412m.lvc}{S190412m} & 24.1 & BBH:$>99$\% & $\sim50$\% & 3.1 \\
\href{https://gcn.gsfc.nasa.gov/notices_l/S190425z.lvc}{S190425z} & 1.3 & BNS:$>99$\% & $\sim2$\% & 0.9 \\
\href{https://gcn.gsfc.nasa.gov/notices_l/S190426c.lvc}{S190426c} & 17.6 & NSBH:60\%, MG: 25\%, BNS:15\% & $\sim20$\% & 2.5 \\
\href{https://gcn.gsfc.nasa.gov/notices_l/S190707q.lvc}{S190707q} & 20.3 &  BBH:$>99$\% & $\sim30$\% & 3.0 \\

\hline
\end{tabular} 
\caption{GW events followed-up with VERITAS so far during the course of the O3 LIGO/Virgo run.}
\label{tab_gwevents}
\end{table*}

The current follow-up strategy involves calculating the visibility of the GW localization region (i.e.\ the amount of observing time available during the current or upcoming night) to select an observable portion. Using a Greedy Traveling Salesman heuristic, the pointings that tile the observable portion are defined and sorted such that the probability covered is maximized and the slewing time is minimized. The exposure per pointing is currently set at 5 min.

An approach is currently being studied that will use both the sky map and distance estimate provided by LIGO/Virgo and a galaxy catalog to select likely hosts of the GW event (similar to that already in use in H.E.S.S.~\cite{Seglar-Arroyo:2017jgz}).

\subsection{Follow-up of realtime and archival single neutrino events from IceCube}

The main component of the neutrino follow-up program is the observation of muon neutrino event positions as reported by IceCube. Only IceCube events from charged-current interactions of muon neutrinos are selected as their typical angular uncertainties ($1^{\circ}$ or better) can be covered by the field of view of the VERITAS cameras. 

This program~\cite{Santander:2015kka} started with the VERITAS observation of neutrino event positions identified by IceCube as potentially astrophysical in origin and reported in publications, which introduced a significant delay between the neutrino detection and the VERITAS follow-up. Since 2016, IceCube started broadcasting the position of likely-astrophysical neutrino events with a latency of at most a few minutes. This enabled prompt follow-up observations by VERITAS and other IACTs~\cite{Santander:2017zkl}.

\begin{figure}
    \centering
    \includegraphics[width=0.5\textwidth]{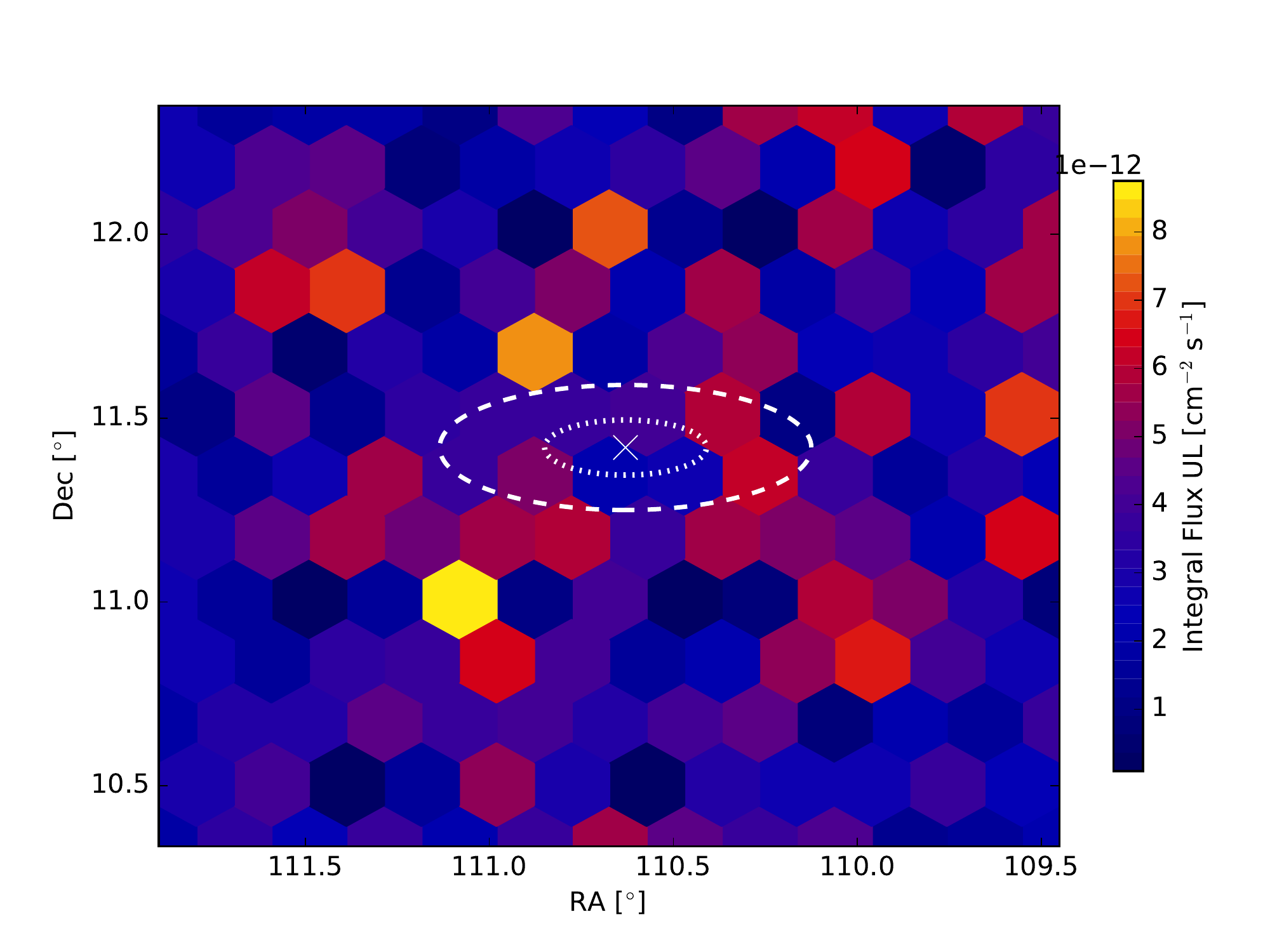}
    \caption{Integral flux upper limit sky map for the PeV muon neutrino event \#27 listed in Ref.~\cite{Aartsen:2016xlq}, for $E > 170$ GeV and assuming a power-law spectral index of $\Gamma=2.5$. The contours indicate the 50\% and 90\% uncertainty region of the neutrino event.}
    \label{fig:icnu20}
\end{figure}

To date, VERITAS has collected over 100 hours of observations following five realtime alerts and 30 archival IceCube neutrinos, shown in the map in Fig.~\ref{fig:allsky}. Deep observations of archival events, such as the PeV muon neutrino event reported in Ref.~\cite{Aartsen:2016xlq}, have been performed by VERITAS. No VHE gamma-ray source was detected in these observations and therefore integral flux upper limits were calculated that cover the neutrino uncertainty region, as shown in Fig.~\ref{fig:icnu20}.

The most interesting follow-up so far was the one performed on the event IC-170922A in 2017, discussed in Section~\ref{subsec:txs}, which resulted in the identification of the first potential astrophysical counterpart to an IceCube neutrino event.

\subsection{Follow-up of IceCube neutrino multiplets}

\begin{figure}
    \centering
    \includegraphics[width=0.44\textwidth]{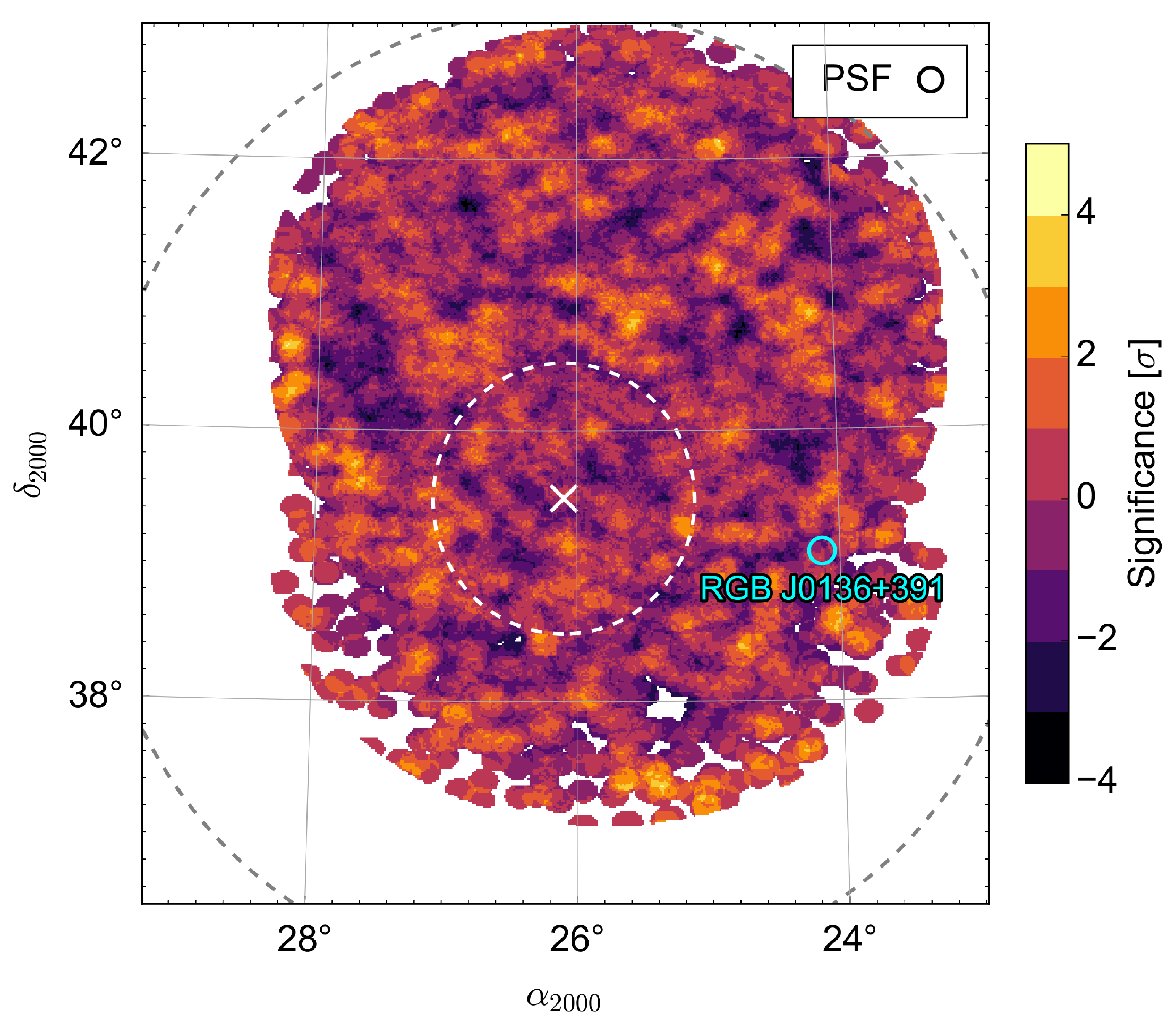}
    \includegraphics[width=0.55\textwidth]{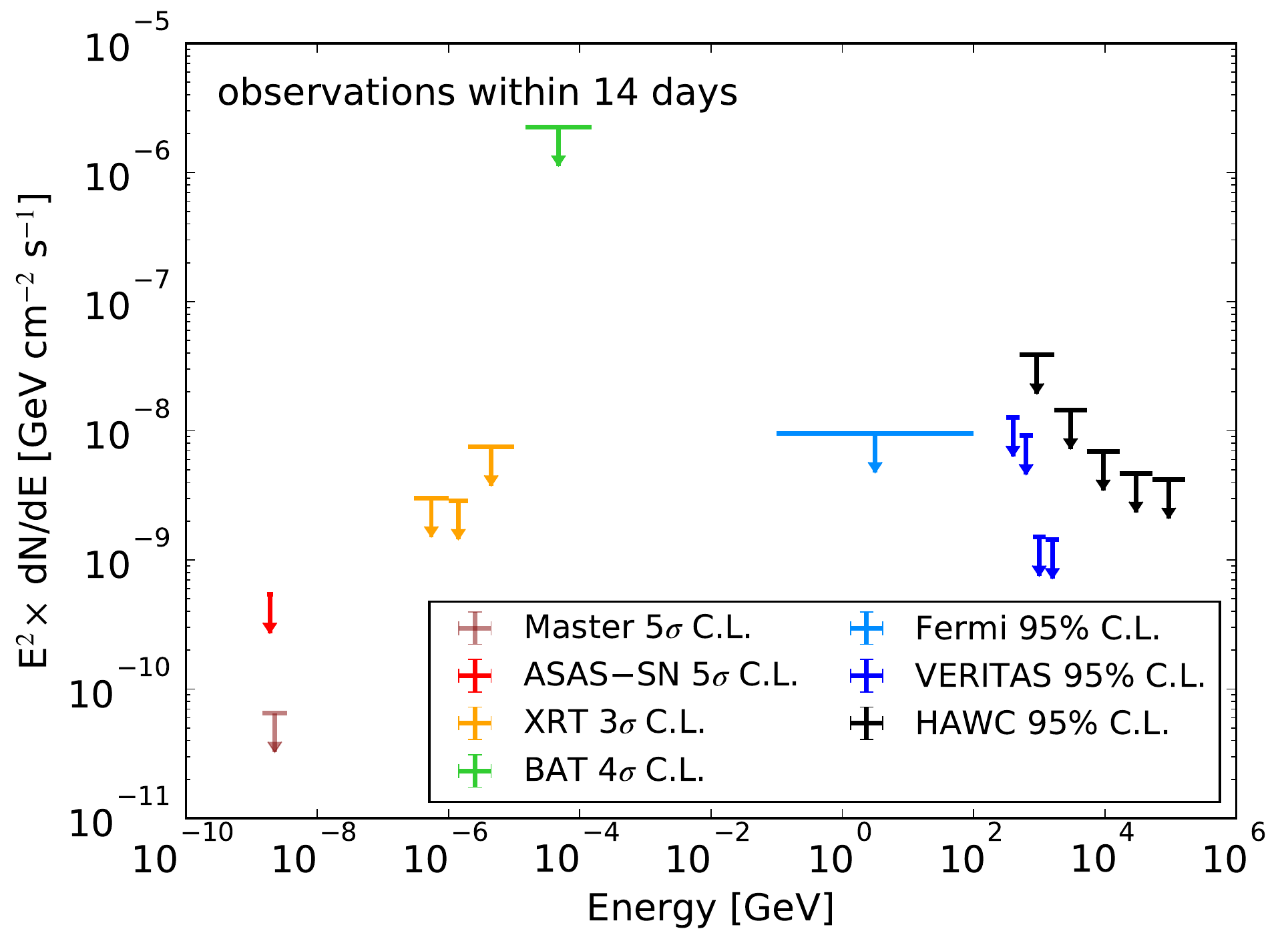}
    \caption{\emph{Left}: Significance sky map indicating the 50\% and 90\% uncertainty region for the triplet (dashed lines) and the location of the closest known VHE source RGB J0136+391. \emph{Right:} Broadband flux upper limits at the triplet central position are shown based on observations collected within 14 days of the triplet detection (from Ref.~\cite{Aartsen:2017snx}).}
    \label{fig:triplet}
\end{figure}

On Feb 17, 2016 IceCube detected a rare neutrino triplet event~\cite{Aartsen:2017snx}. VERITAS received a private notification of the event. Prompt observations of the triplet position were prevented by the full Moon and started on Feb 25. Additional observations were collected on Feb 26 for an effective exposure of 63 min. No new VHE source was detected in the region of the neutrino triplet event. In addition, no signal from the known VHE blazar RGB J0136+391 was found in the VERITAS data. The sky map and flux upper limits are shown in Fig.~\ref{fig:triplet}.

\subsection{Follow-up of potential neutrino emission from known gamma-ray sources}

The major IACTs operate a VHE follow-up program in collaboration with IceCube~\cite{Aartsen:2016qbu}. The IACTs are alerted when a cluster of neutrinos is detected around sources of interest over a certain time period. VERITAS has performed three follow-up observations on blazar targets so far, as shown in the map in Fig.~\ref{fig:allsky}. There has been no detection of VHE activity in these observations that could potentially be connected with the neutrino clusters identified by IceCube.

The source list used in VERITAS to trigger this program has been recently updated with the latest \emph{Fermi}-LAT catalogs to include new sources and optimize the probability of a detection in the area of the sky where IceCube has better sensitivity (i.e.\ near the celestial equator). 

\subsection{Long-term monitoring of the candidate neutrino blazar TXS 0506+056}\label{subsec:txs}

On Sept 22, 2017 IceCube detected an neutrino event candidate (IC170922A) with a high-probability of being astrophysical in origin. Observations collected with the \emph{Fermi}-LAT gamma-ray telescope and multiwavelength observatories identified the flaring blazar TXS 0506+056 as the potential counterpart of IC170922A~\cite{IceCube:2018dnn}. The significance of this correlation is at the $3\sigma$ level. An analysis of archival IceCube data revealed further evidence of historical neutrino emission from the direction of TXS 0506+056~\cite{IceCube:2018cha}, making it compelling target for VHE observations. 

\begin{figure}
    \centering
    \includegraphics[width=0.44\textwidth]{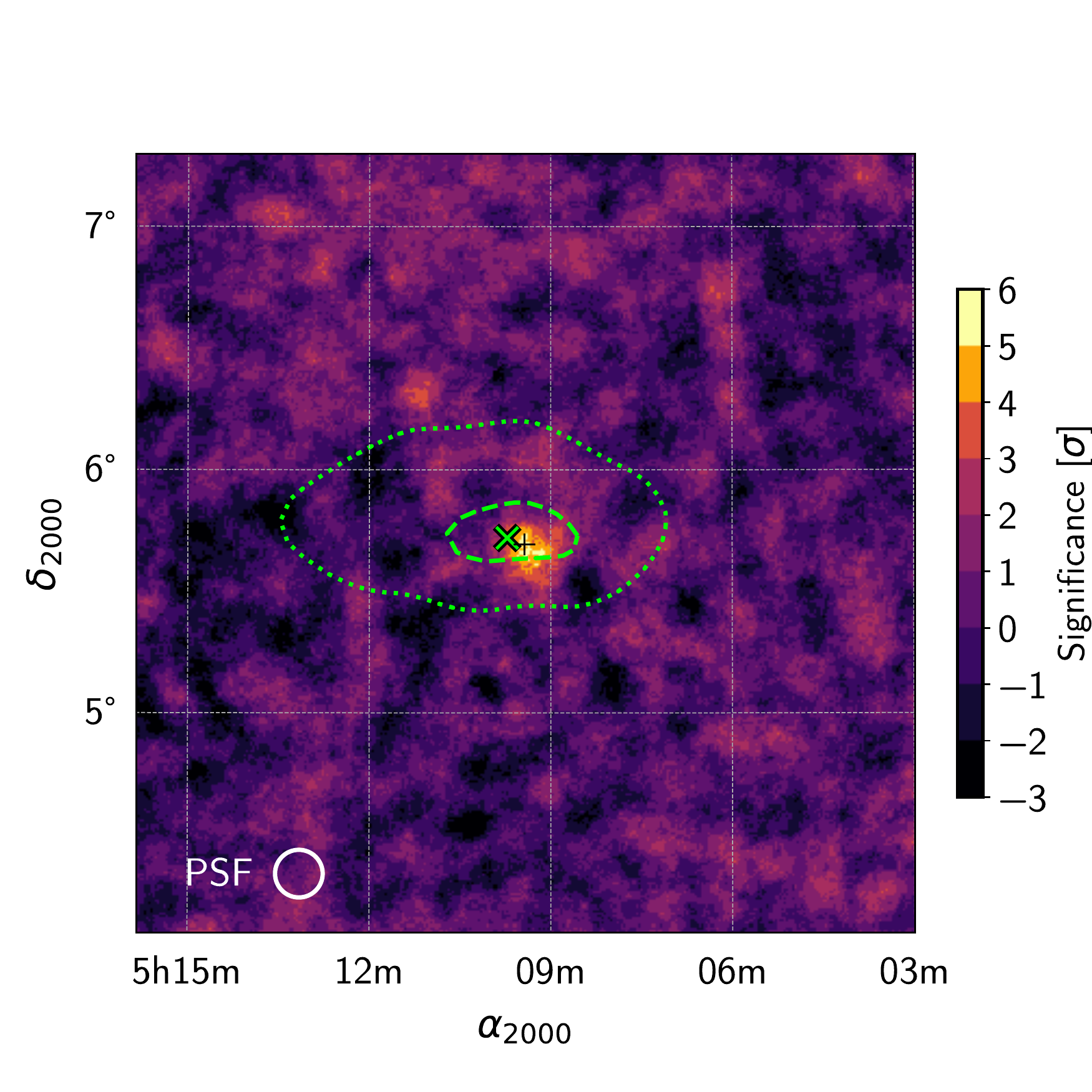}
    \includegraphics[width=0.55\textwidth]{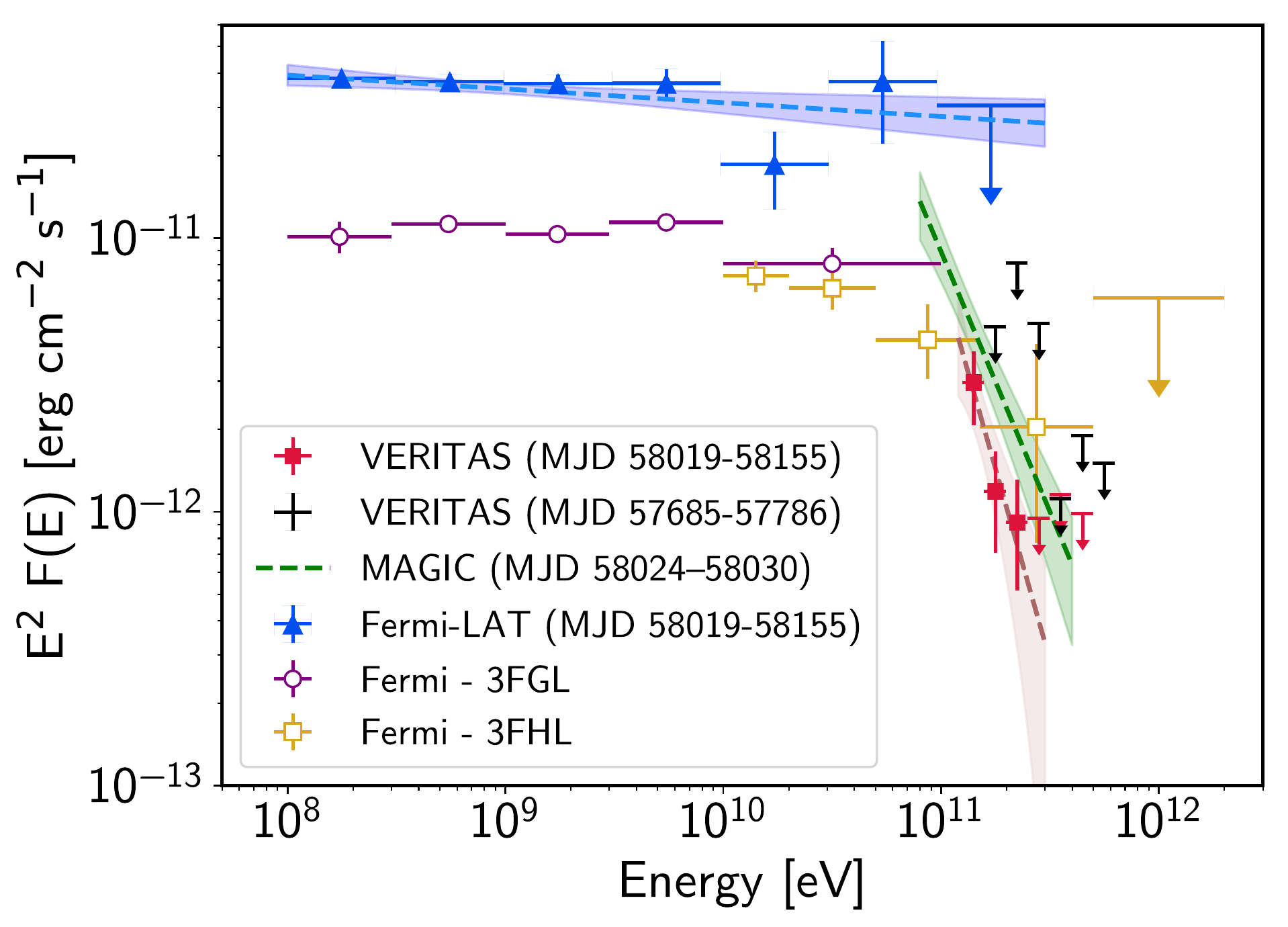}
    \caption{\emph{Left}: VERITAS significance sky map for the location of IC170922A/TXS 0506+056. The green contours indicate the 50\% and 90\% uncertainty region for the neutrino event and the black `+' marker shows the location of the blazar. \emph{Right:} Gamma-ray spectral energy distribution for TXS 0506+056 comparing the soft spectrum measured by VERITAS and MAGIC to the hard spectrum ($\sim E^{-2}$) in the GeV range as measured by \emph{Fermi}-LAT. (See Ref.~\cite{Abeysekara:2018oub} for more details.)}
    \label{fig:txs}
\end{figure}

The source was detected in the VHE range by MAGIC~\cite{Ahnen:2018mvi}, and then by VERITAS~\cite{Abeysekara:2018oub} in extended observations between on September 23, 2017 and February 6, 2018. During this period the source was detected at a significance of 5.8$\sigma$ (see Fig.\ref{fig:txs}) with an average photon flux of (8.9 $\pm$ 1.6) $\times$ 10$^{-12}$ cm$^{-2}$ s$^{-1}$, or 1.6\% of the Crab Nebula flux above 110 GeV. The source displayed a very soft spectral index of $4.8\pm1.3$.

Over the coming years VERITAS will monitor this source to search for additional flares and to better constrain its variability and VHE spectrum. These observations will provide insights into the emission mechanisms at work in the source which can constrain hadronic acceleration models.

\section{Conclusions and outlook}

We have presented a status update on the multimessenger program of the VERITAS observatory, which focuses on the follow-up of neutrino and gravitational wave events. These studies are a central part of the science program of VERITAS. Additional observations will be collected over the coming years to help identify or constrain the VHE gamma-ray emission from multimessenger sources.

\section*{Acknowledgements}

This research is supported by grants from the U.S. Department of Energy Office of Science, the U.S. National Science Foundation and the Smithsonian Institution, and by NSERC in Canada. This research used resources provided by the Open Science Grid, which is supported by the National Science Foundation and the U.S. Department of Energy's Office of Science, and resources of the National Energy Research Scientific Computing Center (NERSC), a U.S. Department of Energy Office of Science User Facility operated under Contract No. DE-AC02-05CH11231. We acknowledge the excellent work of the technical support staff at the Fred Lawrence Whipple Observatory and at the collaborating institutions in the construction and operation of the instrument.

\providecommand{\href}[2]{#2}\begingroup\raggedright\endgroup

\end{document}